# Nonlinear theory of quantum Brownian motion

Roumen Tsekov
DWI, RWTH, 52056 Aachen, Germany

A nonlinear theory of quantum Brownian motion in classical environment is developed based on a thermodynamically enhanced nonlinear Schrödinger equation. The latter is transformed via the Madelung transformation into a nonlinear quantum Smoluchowski-like equation, which is proven to reproduce key results from quantum and classical physics. The application of the theory to a free quantum Brownian particle results in a nonlinear dependence of the position dispersion on time, being quantum generalization of the Einstein law of Brownian motion. It is shown that the time of decoherence for the transition from quantum to classical diffusion is proportional to the square of the thermal de Broglie wavelength divided by the Einstein diffusion constant.

Brownian motion is the permanent irregular movement of a particle immersed in a medium. Its rigorous description requires joint consideration of the coupled dynamics of the Brownian particle and the medium, usually referred as thermal bath. Fundamental problems appear in the Brownian motion theory if the quantum effects become important. For instance, at time larger than the classical momentum relaxation time the Brownian motion of a quantum particle in a classical environment is regularly described by the classical Einstein law $\sigma_x^2 = 2Dt$. It was recently demonstrated that this is not true due to the nonlinear nature of the quantum relaxations (Tsekov 2007). The classical diffusion equation is linear owing to the Boltzmann entropy. More general definitions such as the Rényi and Tsallis entropies, however, lead to nonlinear mean-field Fokker-Planck equations (Frank 2005; Chavanis 2006; Schwämmle et al. 2007). The important point here is that the quantum mechanics is linear on the wave function, not on the probability density as the diffusion equation is. Hence, any probabilistic theory of the thermo-quantum relaxation should be nonlinear (Tsekov 2009) and the proposed linear semiclassical Smoluchowski equations (Ankerhold et al. 2001; Coffey et al. 2007) are only linearization around the equilibrium state (Tsekov 2007). As is shown (Tsekov 1995) the effect of a classical thermal bath on a quantum subsystem results in a nonlinear Schrödinger equation. Thus, the important superposition principle of the quantum mechanics is no more applicable, which leads to decoherence (Joos and Zeh 1985; Zurek 2003; Schlosshauer 2007). The latter is usually described via the Caldeira-Leggett (1983) master equation enhanced by minimal extension with a purely quantum position-diffusion operator to satisfy the Lindblad positivity requirements at some initial states (Petruccione and Vacchini 2005; Bellomo et al. 2007; Vacchini and Hornberger 2007). However, the Caldeira-Leggett master equation is linear due to factorization of the initial density matrix and the enhancement above does not alter the time-dependence of the Einstein law but simply increases the diffusion constant by quantum effects. A heuristic nonlinear quantum relaxation theory led also to position-diffusion operator in the phase space, which persists even in the classical Klein-Kramers-like equation (Tsekov 2001).

If the thermal bath consists of infinite number of harmonic oscillators, the entire dynamics can be rigorously transformed into a stochastic generalized Langevin equation (GLE) for the Brownian particle (Ford et al. 1965, 1988; Ullersma 1966; Benguria and Kac 1981; Lindenberg and West 1984). GLE can be derived in the general case via Mori (1965), Zwanzig (1961) or Lee (1983) formalisms. The main advantage of GLE is a relationship between the memory kernel and the autocorrelation function of the stochastic Langevin force known as the fluctuation-dissipation theorem. GLE belongs to the class of linear response theories as well. They account rigorously for the particle-bath Heisenberg dynamics averaged, however, over an initial non-perturbed bath distribution. Thus, the system and bath are statistically separable, which is similar to the Caldeira-Leggett model. Their quantum levels remain the same as in the isolated systems and only the occupation of states changes in time due to their interaction. This is evident from the fact that the Brownian particle Planck constant does not appear in the Langevin force spectral density, being the same for quantum and classical Brownian particles. To prove this one can employ the method of quantum dynamics with two Planck constants (Diosi 1995). Hence, the present case of a classical thermal bath covers also the well-known high-temperature approximation, where the Langevin force is classical and only the Brownian particle is quantum. Let us examine the range of validity of a quantum linear response theory. If a quantum particle is immersed in an equilibrium thermal bath, the caused perturbation can be regarded as a small one if the particle kinetic energy is lower than the bath thermal energy, i.e. if $\sigma_p^2 \leq m k_B T$, where $\sigma_p^2$ is the particle momentum dispersion. On the other hand, the Heisenberg principle requires $\sigma_x \sigma_p \geq \hbar/2$. Thus, combining these two inequalities it follows that the linear response theory is correct for position deviations $\sigma_x \geq \lambda_T$ larger than the thermal de Broglie wavelength $\lambda_T \equiv \hbar/2\sqrt{m k_B T}$. However, this inequality corresponds to negligible quantum effects on the particle translation. Therefore, a free quantum Brownian particle described by a linear response theory behaves always as a classical one and that is why the Einstein law holds. This is not surprising since the energy spectrum of a free particle in vacuum is continuous. It points out, however, that the approximation of initially separable Brownian particle and bath distributions is rough. The complete description of the quantum Brownian motion requires a nonlinear response theory, which accounts also for the changes of the particle energy levels due to the local force field of the bath particles and vice versa.

The Einstein law $\sigma_x^2 = 2Dt$ implies also that at any time the Brownian particle momentum dispersion is the equilibrium one $\sigma_p^2 = m k_B T$ (Gardiner 2004). Thus, the Heisenberg uncer-tainty principle is violated at short times $t < \lambda_T^2 / 2D$. Such a possibility of equilibrium in the momentum space and non-equilibrium in the coordinate space is only possible for classical particles, since their momentum and coordinate distributions can be factorized. In quantum mechanics these two distributions are interrelated and many problems appear due to the ambiguous interpretation of the Wigner function as probability density. The corresponding Fokker-Planck-like equations allow quicker relaxation of the momentum than that of the coordinate, which will violate the Heisenberg principle at some times. In this case the Wigner function becomes negative, an unacceptable property for a distribution density. A case, where the Einstein law holds for a quantum particle as well, is when the Brownian particle is at equilibrium. In this case, at any time $\sigma_x^2 = \infty$ but one can calculate its quasi-static rate of change. The momentum distributions of a quantum particle at equilibrium and non-equilibrium conditions, however, are always different. Thus, the time-dependent diffusion coefficient calculated from the quantum Smoluchowski equation (Tsekov 2009) will not coincide with that one obtained from the equilibrium Green-Kubo formula (Ford and O'Connell 2006). The latter predicts $\Delta \sigma_x^2 = 2D\Delta t$ in the case of a quantum particle moving in a classical bath since the equilibrium momentum distribution of a free quantum particle is the classical Maxwell one. Let us briefly explain the problem above by giving an example. Imagine at $t = -\infty$ there is a bath coupled to a Brownian particle being in a zero thick and infinitely deep potential well. Nevertheless that the system is equilibrated, one knows exactly that the Brownian particle is in the potential well at any time. At time $t = 0$ the external potential well is switched off and the Brownian particle is free to diffuse now. If the Brownian particle is a classical one the removal of the potential well will not disturb its momentum distribution. On the contrary, if the Brownian particle is a quantum one, after removal of the external potential its initial momentum distribution will not be the equilibrium one following at $t = \infty$. Initially the particle will possess infinite momentum dispersion and it will take some time until it is dissipated in the bath to reach the equilibrium value $\sigma_p^2 = mk_BT$. Since the Brownian particle and bath are permanently coupled (Tsekov and Ruckenstein 1994), the switch-off of the external potential will affect also the bath particles. And because the potential well possesses infinite energy its removal cannot be described by a linear response theory applicable to perturbations with typical energy of the order of the thermal one. The problem can be solved via an evolution equation for the probability density with delta-function initial distribution density, which is the aim of the following theory.

In the present paper, our original approach (Tsekov 1995) based on thermodynamic extension of the Schrödinger equation is refined and further developed to describe properly the inertial effects as well. The statistical correlations between the Brownian and bath particles are accounted for thus leading naturally to dissipation of energy. Let us start most generally with the Schrödinger equation for the coupled system Brownian particle and bath

$$i\hbar\partial_t\psi = (\hat{p}^2/2m + U + W + \hat{H}_B)\psi \qquad (1)$$

which will never be considered as separable in the present theory. Here $\psi$ is the system wave function, $\hat{p} \equiv -i\hbar\nabla$ is the momentum operator of a Brownian particle with mass $m$, $U(r)$ is an external potential acting on the Brownian particle only, $W(r,R)$ is a potential accounting for the particle-bath interaction and $\hat{H}_B$ is the pure bath Hamiltonian. The statistical nature of the wave function implies the following Bayesian decomposition, $\psi = \Phi\phi$, where $\phi(r,t)$ is the wave function of the Brownian particle and $\Phi(R,t|r)$ is the conditional wave function of the bath. The square on the latter represents the conditional probability to find a specific configuration $R$ of the bath particles at time $t$ if at this moment the Brownian particle occupies the point $r$. It is important to note that $\phi$ and $\Phi$ are wave functions, i.e. they represent probability densities, which can be calculated directly from $\psi$. Thus, the decomposition above is not arbitrary and it should preserve the probability and momentum of the whole system. Substituting $\psi = \Phi\phi$ in Eq. (1), multiplying the result by the complex-conjugated bath wave function and integrating over the bath particles coordinates yields

$$i\hbar\partial_t\phi = (\hat{p}^2/2m + U + \langle\Phi|W|\Phi\rangle + \langle\Phi|\hat{H}_B - i\hbar\partial_t|\Phi\rangle + \langle\Phi|\hat{p}\Phi\rangle\cdot\hat{p}/m + \langle\Phi|\hat{p}^2\Phi\rangle/2m)\phi \qquad (2)$$

The integral $\langle\Phi|W|\Phi\rangle \equiv \overline{W}$ represents the mean potential of the Brownian particle-bath interaction. If the bath is a solid body $\overline{W}$ is a periodic function on $r$, while a constant $\overline{W}$ is a good model for a fluid bath (Tsekov and Ruckenstein 1994). The term $\langle\Phi|i\hbar\partial_t - \hat{H}_B|\Phi\rangle \equiv TS$ accounts obviously for energy fluctuations solely related to the bath probability density changes caused by the Brownian particle. Thus, it is attributed to the entropy $S$ of the Brownian particle. The integral $\langle\Phi|\hat{p}\Phi\rangle$ is real since $\Phi$ is normalized. Note that the operator $\hat{p}$ acts on the conditional variable $r$, not on $R$. This integral has the meaning of momentum loss of the Brownian particle via friction. For the Ohmic resistance it can be described as $\langle\Phi|\hat{p}\Phi\rangle = -br$ in analogy with the classical expression $\Delta p = -b\int \dot{r}dt$, where $b$ is the friction coefficient of the Brownian particle, which could depend on temperature $T$. Applying the momentum operator $\hat{p}$ on this relation again yields an expression for the last term in Eq. (2)

$$\langle\Phi|\hat{p}^2\Phi\rangle/2m = 3i\hbar b/2m - \langle\hat{p}\Phi|\hat{p}\Phi\rangle/2m \qquad (3)$$

The integral on the right hand side is also a real function and represents the heat emitted by the Brownian particle in the bath via friction. $\langle \hat{p}\Phi | \hat{p}\Phi \rangle / 2m = -bA/m$ is proportional to the action $A \equiv \int p \cdot dr$ and friction coefficient $b$ in accordance to the classical expression $\Delta p^2 / 2m = -b\int \dot{r}^2 dt$ following from the Rayleigh dissipative function. Substituting all these integrals in Eq. (2), the latter acquires the form

$$i\hbar \partial_t \phi = [\hat{p}^2 / 2m + U + \overline{W} - TS + bA/m - b(\hat{p} \cdot r + r \cdot \hat{p})/2m]\phi \qquad (4)$$

Since the entropy $S$ and the action $A$ depend on the particle wave function $\phi$, Eq. (4) is a nonlinear Schrödinger equation (Pang and Feng 2005). Hence, the superposition principle is not valid anymore and the energy levels of the Brownian particle permanently change in time. The entropic term in Eq. (4) represents an original thermodynamic DFT potential. One can model specific statistical properties of the Brownian and bath particles via $S(\rho)$, where $\rho \equiv \phi^*\phi$ is the probability density. For instance, the Gross-Pitaevskii BEC theory is based on the cubic Schrödinger equation with a linear entropy $S \sim \rho$. The Boltzmann entropy $S = -k_B \ln \rho$ leads to the logarithmic Schrödinger equation (Bialynicki-Birula and Mycielski 1976, 1979; Davidson 1979, 2001). Note that $S$ is the local entropy and the Gibbs definition refers to its mean value $\int S\rho dr$. While the entropy causes thermodynamic decoherence disappearing at zero temperature, the action $A$ induces frictional decoherence. The action term $bA/m$ in Eq. (4) is identical to the frictional term $i\hbar b \ln(\phi^*/\phi)/2m$ in the Schrödinger-Langevin equation (Kostin 1972, 1975; Davidson 1979). A logarithmic Schrödinger-Langevin equation is also derived (Nassar 1985). Finally, chemical reactions between the Brownian and bath particles can be accounted for by an effective DFT potential $\overline{W}(\rho)$ (Ivanov 1980). In fact, the term $\overline{W} - TS$ represents the Brownian particle chemical potential, i.e. the work necessary to remove quasi-statically the quantum Brownian particle from the bath at constant volume and temperature. The further applications of the nonlinear Schrödinger equation (4) require specification of the entropy. A transparent way to model $S$ occurs when the complex wave function $\phi$ is expressed by $\rho$ and $A$ via the well-known Madelung transformation $\phi = \sqrt{\rho}\exp(iA/\hbar)$. Substituting it in Eq. (4) leads to the following two equations corresponding to the imaginary and real parts, respectively,

$$\partial_t \rho + \nabla \cdot (\rho V) = 0 \qquad m\partial_t V + mV \cdot \nabla V + bV = -\nabla(Q + U + \overline{W} - TS) \qquad (5)$$

where the velocity in the probability space is defined by $V \equiv (\nabla A - br)/m$. These equations are similar to the Madelung (1927) quantum hydrodynamics and the Bohmian (1952) mechanics, where the quantum potential $Q \equiv -\hbar^2 \nabla^2 \sqrt{\rho} / 2m\sqrt{\rho}$ (Carroll 2007) accounts for all quantum effects. Note that in the classical limit $Q$ disappears and Eqs. (5) reduce to those of the classical Brownian motion theory (Gardiner 2004). It points out also the shortcoming of the linear response theory, where the probability density should be replaced in $Q$ by its equilibrium expression to linearize Eqs. (5). Since the equilibrium density for a free quantum Brownian particle is uniform, the quantum potential vanishes and thus the linear theory description coincides with the classical limit (Tsekov 2007).

Let us consider first the case of zero temperature, where the entropic term in Eqs. (5) drops out. Thus, the particle-bath interaction is due only to scattering of the moving Brownian particle on the static but movable bath particles. If one considers a constant force field $U + \overline{W} = -fx$, the solutions of Eqs. (5) at $T = 0$ are

$$\rho = \exp[-(x-\mu)^2/2\sigma_x^2]/\sqrt{2\pi\sigma_x^2} \qquad V = \partial_t \mu + (x-\mu)\partial_t \sigma_x / \sigma_x \qquad (6)$$

The mean value $\mu$ and the dispersion $\sigma_x^2$ obey the following dynamic equations

$$m\partial_t^2 \mu + b\partial_t \mu = f \qquad m\partial_t^2 \sigma_x + b\partial_t \sigma_x = \hbar^2 / 4m\sigma_x^3 \qquad (7)$$

The first one reflects the Ehrenfest theorem. The second equation is nonlinear due to the quantum potential. However, there are two simple quantum limits. In the case of a single particle in vacuum $(b=0)$ the solution of Eq. (7) is the known expression $\sigma_x^2 = \sigma_x^2(0) + [\hbar t / 2m\sigma_x(0)]^2$ for the spreading of a Gaussian wave packet with $\sigma_p = \hbar / 2\sigma_x(0)$. Considering now the hypothetical experiment with the potential well discussed before, initially $\sigma_x^2(0) = 0$ but at any other time the wave packet is already spread $\sigma_x^2(t>0) = \infty$ due to the constantly infinite momentum dispersion. In the opposite case of strong friction, the first inertial term in Eq. (7) is negligible as compared to the second one and the solution reads $\sigma_x^4 = \hbar^2 t / mb$. The proportionality $\sigma_x \sim \sqrt[4]{t}$ is detected in numerical simulations of electrons as a limit at vanishing hopping strength corresponding to $T \to 0$. In general, $\sigma_x \sim t^\alpha$ and the numerical simulations (Cerovski et al. 2005) show linear dependence of $\alpha$ on the hopping strength, spanned between 0.25 and the ballistic 1 (the Einstein law corresponds to $\alpha = 0.5$). As seen, in the high friction limit $\sigma_x^2$ is not proportional to $t^2$ since the particle is not in vacuum anymore. The bath not simply decreases the kinetic energy of the particle but changes its energy spectrum as well via the harmonic potential

$bA/m$ in Eq. (4). At $t=0$ again $\sigma_x^2 = 0$ and $\sigma_p^2 = \infty$, but with increasing time the particle loses its momentum dispersion $\sigma_p^2 = \hbar\sqrt{mb/t}/4$ due to the friction. This is compensated, however, by increase of the position dispersion $\sigma_x^2 = \hbar\sqrt{t/mb}$ to satisfy the minimal Heisenberg principle. Note that in the limit case $b \to \infty$ the initial state will last forever, since the particle cannot move. In fact, the Brownian particle is trapped now in a kinetic well. In the quantum case, this will restrict the relaxation of the momentum dispersion as well due to the Heisenberg principle. According to the linear response theory, $\sigma_p^2$ will drop to zero immediately due to the zero classical momentum relaxation time $m/b \to 0$ thus violating the Heisenberg principle since $\sigma_x^2 = 0$ as well. The reason that this does not happen in reality is that according to Eq. (4) the distance between the particle energy levels tends also to infinity at $b \to \infty$.

Equations (5) are heuristically proposed by Tsekov and Vayssilov (1992). Nevertheless, that the attempt to specify the chemical potential solely on the equilibrium distribution fails, at $T=0$ the obtained equation coincides with Eq. (7). The problem is that in general $T$ is the local non-equilibrium temperature in Eqs. (5). It is introduced in Eq. (4) as a parameter, reflecting the complex chaotic motion of the bath particles. A very fast Brownian particle will obviously destroy completely the local thermodynamic equilibrium. Hence, one has to restrict the consideration to relatively low velocities in order to be able to employ the equilibrium statistical temperature, which appears in the linear response theory via the initial bath distribution. Thus, one should linearize Eqs. (5) on $V$ to obtain

$$m\partial_t^2 \rho = \nabla \cdot [b\rho V + \rho \nabla (Q + U + \overline{W} - TS) + m\nabla \cdot (\rho VV)] = -b\partial_t \rho + \nabla \cdot (\rho \nabla F) + O(V^2) \qquad (8)$$

where the local free energy potential is introduced via the relation $F \equiv Q + U + \overline{W} - TS$. Since the thermodynamic entropy $S \equiv -(\partial_T F)_{\rho,b}$ is the temperature derivative of the free energy, one can express $F$ by integration on $T$ as follows

$$F + TS_{\beta=0} = k_B T \int_0^\beta (Q + U + \overline{W})_b d\beta + k_B T \ln(\rho/\rho_{\beta=0}) = k_B T [\int_0^\beta \frac{1}{\sqrt{\rho}} (\hat{H} + 2\partial_\beta)\sqrt{\rho} d\beta]_b \qquad (9)$$

where the logarithm of the probability density appears as integration constant and represents the Boltzmann entropy, $\beta \equiv 1/k_B T$ and $\hat{H} \equiv \hat{p}^2/2m + U + \overline{W}$ is the Brownian particle Hamiltonian. The subscript $_b$ indicates that in this thermodynamic relation the friction coefficient $b$ must be considered constant during the integration on $\beta$. Note that the entropy possesses non-Boltzmannian components as well due to the temperature dependence of the average $\overline{W}$ and quantum $Q$ potentials. Thus, the latter leads to the following quantum entropy

$$S_Q \equiv (Q - k_B T \int_0^\beta Q d\beta)/T \tag{10}$$

which could probably explain the observed differences between the thermodynamic and von Neumann entropy definitions at low temperature (Hörhammer and Büttner 2008). Since at infinite temperature the local entropy is uniform the term $TS_{\beta=0}$ is constant not affecting the Brownian motion. Substituting Eq. (9) in Eq. (8) the following nonlinear differential equation is obtained

$$m\partial_t^2 \rho + b\partial_t \rho = k_B T \nabla \cdot [\rho \nabla \int_0^\beta \frac{1}{\sqrt{\rho}} (\hat{H} + 2\partial_\beta)\sqrt{\rho} d\beta]_b \tag{11}$$

which describes the quantum Brownian motion beyond the linear response. In the classical limit Eq. (11) reduces at $\overline{W} = 0$ to the well-known telegraph-like equation (Gardiner 2004)

$$m\partial_t^2 \rho + b\partial_t \rho = \nabla \cdot (\rho \nabla U + k_B T \nabla \rho) \tag{12}$$

which is linear. If the first inertial term is neglected, Eq. (11) converts into a nonlinear quantum Smoluchowski equation (Tsekov 1995). Note that the nonlinear quantum-diffusive structure is universal and holds in any representation of quantum mechanics (Tsekov 2001). At zero temperature Eq. (11) reduces to the following nonlinear equation

$$m\partial_t^2 \rho + b\partial_t \rho = \nabla \cdot [\rho \nabla (Q + U + \overline{W})] \tag{13}$$

describing purely quantum diffusion in the field of an external potential. As seen, the quantum potential drives the quantum diffusion, which particularly manifests itself into the tunneling effect. In the case of a free Brownian particle the distribution density is Gaussian and Eq. (13) can be rewritten in the usual telegraph form $m\partial_t^2 \rho + b\partial_t \rho = (\hbar^2/4m\sigma_x^2)\nabla^2 \rho$. Hence, the quantum diffusion is driven by the minimal Heisenberg momentum uncertainty. Due to the neglected quadratic velocity terms in Eq. (8), this equation reproduces the results from Eq. (7) at relatively high friction constant only. In the case of strong friction it simplifies further to a diffusion equation (Tsekov 2009) with a dispersion-dependent quantum diffusion coefficient $D_Q \equiv \hbar^2/4mb\sigma_x^2$.

At equilibrium, the probability density does not depend on time and the probability flux is zero. Therefore, according to Eq. (11) the equilibrium probability density $\rho_e$ obeys the following equation

$$-2\partial_\beta \sqrt{Z\rho_e} = \hat{H}\sqrt{Z\rho_e} \qquad (14)$$

where $Z$ is the normalization factor. Equation (14) is a Schrödinger one, where the time is replaced by the imaginary time $\beta\hbar/2i$, i.e. a half Wick rotation (Kostin 1995). If the potential $\overline{W}$ is temperature independent, the solutions of Eq. (14) are $\{\rho_n = Z^{-1}\exp(-\beta E_n)\phi_n^2\}$, where $\{E_n, \phi_n\}$ are the eigenvalues and orthonormal real eigenfunctions of the Brownian particle Hamiltonian, $\hat{H}\phi_n = E_n\phi_n$. As expected, the quantum particle is described not only by coordinate but also by its quantum state. The probability density above is product of the spatial distribution in a given quantum state and the canonical Gibbs probability for occupation of this state at a given temperature. From the normalization it follows that $Z$ is the canonical partition function $\sum \exp(-\beta E_n)$. Equation (14) has the following formal solution (for simplicity $\overline{W} = 0$)

$$\rho_e = Z^{-1}\exp[-\beta U - \int_0^\beta Q(\rho_e)d\beta] = Z_{sc}^{-1}\exp\{-\beta U - \beta^2\hbar^2[3\Delta U - \beta(\nabla U)^2]/24m\} \qquad (15)$$

where the last semiclassical approximation is obtained via substitution of $\rho_e$ in the quantum potential $Q$ by the classical Boltzmann distribution $\rho_{cl} = Z_{cl}^{-1}\exp(-\beta U)$. The semiclassical distribution (15) coincides with one derived by the Feynman-Vernon path integrals (Ankerhold et al. 2005) and contradicts to another one derived by the Wigner function (Coffey et al. 2007). This is not surprising, however, since thermodynamic studies have also shown that from the Wigner-Kirkwood and Feynman-Hibbs effective potentials only the latter agrees quantitatively with quantum simulations (Calvo et al. 2001). Applying the same linearization procedure to the quantum potential in Eq. (11) results in the following linear semiclassical telegraph-like equation

$$m\partial_t^2\rho + b\partial_t\rho = \nabla\cdot(\rho\nabla U_{eff} + k_B T\nabla\rho) \qquad (16)$$

where $U_{eff} \equiv U + \beta\hbar^2[3\Delta U - \beta(\nabla U)^2]/24m$ is an effective potential accounting solely for all quantum effects, while the diffusion constant remains classical. If additionally one neglects the first inertial term and employs the following approximation being valid close to the equilibrium, $\rho\nabla U_{eff} \approx \rho\nabla(U + \beta\hbar^2\Delta U/24m) + \beta\hbar^2\nabla\cdot(\rho\nabla\nabla U)/12m$, Eq. (16) reduces to an equation with position-dependent diffusion coefficient (Ankerhold et al. 2001, 2005). Naturally, the equilibrium solution of Eq. (16) is the distribution (15). Due to the linearization around the equilibrium Boltzmann distribution, Eq. (16) corresponds to relaxed quantum fluctuations and if the external potential is omitted it reduces to the classical telegraph equation (12). Note that the rigor-ous linearization of Eq. (11) is complex nontrivial mathematical problem, which certainly will reflect in additional higher-order differential operators acting on the probability density.

It is difficult to solve in general the non-equilibrium equation (11) but if the acting potential is a harmonic one, $U + \overline{W} = m\omega_0^2 x^2/2 - fx$, the Gaussian distribution (6) is again the solution. The corresponding mean value and dispersion obey the dynamic equations

$$m\partial_t^2 \mu + b\partial_t \mu + m\omega_0^2 \mu = f \qquad m\partial_t^2 \sigma_x^2 + b\partial_t \sigma_x^2 + 2m(\omega_0^2 - k_B T \int_0^\beta \frac{\hbar^2}{4m^2 \sigma_x^4} d\beta)_b \sigma_x^2 = 2k_B T \qquad (17)$$

The first equation represents the Ehrenfest theorem again. As seen, the quantum effect in the second equation simply reduces the oscillator spring constant. This is expected, however, since for Gaussian distributions the quantum potential is harmonic. Due to mathematical difficulties it is impossible to solve Eq. (17) explicitly but some well-known results are recovered. For instance, the equilibrium dispersion $\sigma_x^2 = (\hbar/2m\omega_0)\coth(\beta\hbar\omega_0/2)$ coincides with the result from the statistical thermodynamics. Let us consider now the most interesting case of a free Brownian particle ($\omega_0 = 0$) in the high friction limit relevant to the Einstein law. In this case Eq. (17) reduces to

$$\partial_t \sigma_x^2 = 2D(1 + \sigma_x^2 \int_0^\beta \frac{\hbar^2}{4m\sigma_x^4} d\beta)_b \qquad (18)$$

where $D \equiv k_B T/b$ is the Einstein diffusion constant. The term in the brackets represents the relative increase of the diffusion coefficient due to the quantum potential. In the case $T = 0$ Eq. (18) reduces to the overdamped limit of Eq. (7). Since the dispersion $\sigma_x^2$ increases in time, at large times the quantum term is negligible and the solution of Eq. (18) tends asymptotically to the Einstein law $\sigma_x^2 = 2Dt$. At small $t$ the quantum term is dominant and one can neglect now the unity in the brackets of Eq. (18). The solution of the remaining equation is again the already derived expression for the purely quantum diffusion

$$\sigma_x^2 = \hbar\sqrt{t/mb} \qquad (19)$$

Another simple solution of Eq. (18) is the semiclassical limit $\sigma_x^2 = 2Dt + \lambda_T^2 \ln(2Dt/\lambda_T^2)/3$, which is obtained by replacing of the dispersion in the quantum term with the classical expression $\sigma_x^2 = 2Dt$. Note that it becomes negative at short time but this is a defect of the semiclassical approach since at $t < \lambda_T^2/2D$ the diffusion is predominately quantum. In this case $2Dt$ is the classical correction to Eq. (19) and the following superposition represents, in fact, the semi-quantum solution of Eq. (18)

$$\sigma_x^2 = \hbar\sqrt{t/mb} + 2Dt = 2\sqrt{Dt}(\sqrt{Dt} + \lambda_T) \qquad (20)$$

Due to the different time-dependencies of the classical and quantum diffusions, Eq. (20) possesses correct limits both at short and large times. Thus, it is the first order quantum generalization of the classical Einstein law of Brownian motion. A second order approximation can be obtained by substituting of the superposition (20) on the right hand side of Eq. (18) and integrating of the result on time and temperature. This iterative procedure can be repeated as much as possible.

Another approximation, related now to the temperature dependence of $\sigma_x^2$, can be derived by neglecting the quantum entropic effect. Thus, performing the integration in Eq. (18) at constant $\sigma_x^4$ yields (Tsekov 2007)

$$\partial_t \sigma_x^2 = 2D(1 + \lambda_T^2/\sigma_x^2) = 2(D + D_Q) \qquad (21)$$

which differs from Eq. (7) by neglected inertial term and added thermal momentum dispersion of the Brownian particle. Since $\sigma_x^2$ increases with temperature, the solution of Eq. (21) is the upper bound of the exact solution of Eq. (18). According to the nonlinear equation (21), the relative quantum effect on the effective diffusion coefficient is proportional to the ratio between the square of the thermal de Broglie wavelength $\lambda_T$ and the dispersion $\sigma_x^2$. Hence, the quantum effect vanishes at $\sigma_x \gg \lambda_T$ thus leading to transition from quantum to classical diffusion. If one attempts to linearize Eq. (21) around the equilibrium value $\sigma_x^2 = \infty$, it reduces always to the classical limit $\sigma_x^2 = 2Dt$. This confirms again that the quantum diffusion effect cannot be described by the linear response theory. The integration of Eq. (21) is straightforward and the result is

$$\sigma_x^2 - \lambda_T^2 \ln(1 + \sigma_x^2/\lambda_T^2) = 2Dt \qquad (22)$$

The explicit time-dependence of the dispersion $\sigma_x^2 = \lambda_T^2\{-1 - W_{-1}[-\exp(-1 - 2Dt/\lambda_T^2)]\}$ involves a Lambert *W*-function (Tsekov 2007). This upper limit is better approximation of $\sigma_x^2$, since it is always smaller than the superposition (20). The dispersion depends nonlinearly on time in the beginning, but at large times the dependence is almost linear. A good approximation

$\sigma_x^2 = 2Dt + 2\lambda_T^2 \ln(1 + \sqrt{Dt}/\lambda_T)$ via elementary mathematical functions is obtained by substitution of $\sigma_x^2$ by Eq. (20) in the quantum term of Eq. (21) and integration on time. At short time it reduces to Eq. (20), while at large time the limit $\sigma_x^2 = 2Dt + \lambda_T^2 \ln(Dt/\lambda_T^2)$ is always larger than the exact semiclassical limit due to the neglected quantum entropy. Finally, if $\partial_t \sigma_x^2 \sim \sigma_x^2/t$ one would expect a good approximation of the solution of Eq. (18) in the form

$$\sigma_x^2 = 2\lambda_T \sqrt{Dt} \coth(\lambda_T/\sqrt{Dt}) \tag{23}$$

which provides correct limits at short and long times. It is closer to the exact semiclassical limit.

As seen, the quantum effects are essential in the beginning. The characteristic time $\lambda_T^2/2D$ of dynamic decoherence marks the transition from quantum to classical diffusion. From the inequality $\lambda_T^2/D > m/b$ one can estimate the necessary value of the friction coefficient $b > 2mk_BT/\hbar$ to be able to neglect the inertial but still to observe quantum effects. It corresponds to $D < \hbar/2m$ as well. As was mentioned in the introduction the classical Einstein law violates the Heisenberg principle for times shorter than $\lambda_T^2/2D$ due to $\sigma_p^2 = mk_BT$. According to Eq. (21), the non-equilibrium momentum dispersion is given by the expression $\sigma_p^2 = mk_BT + \hbar^2/4\sigma_x^2$, which is valid in the high friction limit only (Tsekov 2009). It satisfies the Heisenberg principle at any time and reduces at infinite time to the known equilibrium momentum $\sigma_p^2 = mk_BT$ and position $\sigma_x^2 = \infty$ dispersions. Note that the decay of $\sigma_p^2$ in time is non-exponential. This non-equilibrium expression describes spreading of a Gaussian wave packet, which is continuously monitored by the thermal bath with measurement uncertainty $\sigma_x$. This happens via reversible $S$ and irreversible $-bA/mT$ entropy changes, which according to the Shannon theory are equivalent to information exchange (Brasher 1991). Therefore, the measurements are not a privilege of the human being only. They exist in any open system divided to observable subsystem and non-observable interacting environment. Moreover, since the measured information is associated with nonlinear entropic terms in the Schrödinger equation the collapse of the wave function appears naturally in the measurement process. For instance, in the Schrödinger cat paradox the entropy of mixing causes nonlinear terms in the Schrödinger equation and thus the outcome is either alive of dead cat but never a linear superposition of this excluding events (Pearle 1976). Therefore, Eq. (4) is applicable also to living systems, which are generally recognized by negative entropy production. A new important aspect appears here due to the quantum entropy. Since the Bohm quantum potential is not local, in contrast to the Boltzmann entropy $S_Q$ is a non-local quantity. Hence, the quantum entropy could cause distant quantum communications between the system and its environment. An interesting fact is also that the quantum potential itself is proportional to the local Fisher information stored in the probability density (Carroll 2007). Thus, $Q$ possesses a non-thermal entropic origin and that is why it drives the purely quantum diffusion (Tsekov 2008).

The present theory is not a Bohmian mechanics since the Bohmian trajectories, defined via $\dot{R} = V(R,t)$ (Bohm 1952), are not involved. Equations (5) describe solely the probability density evolution and $V$ is the velocity in the probability space. However, it is recently shown (Tsekov 2009) that a stochastic Bohm-Langevin equation could be the background for the thermo-quantum diffusion, thus making a bridge between the thermal Brownian motion and the Bohmian mechanics. Possible generalizations can be explored via the Nelson stochastic mechanics (Davidson 1979; Nassar 1985) to derive Eq. (4) from first principles. According to the present theory the thermo-quantum dynamics of an open system obey a nonlinear Schrödinger equation. The original linear Schrödinger equation (1) is applicable to closed systems only but they are not present in Nature due to the infinite range of the fundamental interactions (the only exception is the whole Universe). This is also evident from Eq. (4), where the frictional and entropic terms could never vanish due to the second and third laws of thermodynamics, respectively. Therefore, the linear Schrödinger equation is an approximation and this questions the precision of many effects due to the superposition principle. For instance, the Heisenberg matrix mechanics, Fourier transform and Wigner function are no more applicable, while the Bohmian mechanics seems more plausible. Another example is the traditional theory of decoherence, which is based on linear master equations (Schlosshauer 2007). Thus, decoherence takes place continuously in time. According to the present theory the entropy (information) exchange between the system and its environment results in a nonlinear Schrödinger equation. Hence, from the very beginning decoherence takes place, thus solving the contradiction between quantum and classical reality immediately.